# "Spin-Hall Effect": From the Ballistic to the Diffusive Regime

Roksana Golizadeh-Mojarad[1] and Supriyo Datta[2]
School of Electrical and Computer Engineering,
and the NSF Network for Computational Nanotechnology
Purdue University, West Lafayette, IN-47906, USA.
(Dated: April 14, 2007)

The "spin-Hall effect" has recently attracted a lot of attention and a central question is whether the effect is due to the *intrinsic* spin-orbit interaction or due to spin-asymmetric scattering by *extrinsic* impurities. The objective of this paper is to shed light on this question using a new approach based on the non-equilibrium Green function (NEGF) formalism, which allows us to go continuously from the ballistic to the diffusive limit and we present approximate analytical expressions that describe our results fairly well. Our model suggests that a spin accumulation proportional to the current should be observed even in clean ballistic samples and we show how this spin accumulation evolves as momentum and/or spin relaxation processes are introduced in a controlled way. We further show good quantitative agreement with recent experimental observations in GaAs suggesting that these can be understood in terms of an intrinsic effect driven by the Rashba interaction, although experiments on ZnSe likely have a different origin.

The "spin-Hall effect"[1] has attracted a lot of attention, especially following recent experimental observations of the effect, both optically[2,3,4] and electronically[5]. One central issue regarding the underlying mechanism is whether the effect is due to the *intrinsic* spin-orbit interaction or due to spin-asymmetric scattering by *extrinsic* impurities and the answer is not yet clear[6]. Theoretically, the question has been addressed largely using the Kubo formalism[7,8], the drift-diffusion formulation[9] and very recently the scattering theory of transport[10,11]. The objective of this paper is to shed light on this question using a very different approach based on the non-equilibrium Green function (NEGF) method[12,13], which allows us to study the evolution of the effect continuously from the ballistic to the diffusive limit. We use quotes around "spin-Hall effect" since this term is usually associated with the existence of a non-zero bulk spin conductivity tensor[7,8], which is calculated from microscopic theory[1] and can then be used in a macroscopic transport theory. We do not calculate this tensor and as such do not offer any insights regarding it. Instead, we use our NEGF-based approach to connect directly from a microscopic Hamiltonian to the macroscopic effects that are actually observed in experiments, namely an accumulation of spin-up and spin-down electrons at the opposite edges of a finite-sized sample. We call this a "spin-Hall effect" simply by its analogy with the normal Hall effect, which involves the accumulation and depletion of charges at opposite edges (see Fig. 1). In either case, the accumulation (spin or charge) vanishes for zero current and is proportional to the current at least for low bias.

*Model:* We consider a two-dimensional conductor described by a single-band effective mass ($m^*$) Hamiltonian

$$H_0 \equiv (p_x^2/2m^*) + (p_y^2/2m^*) + U(y), \quad (1)$$

with a confinement potential $U(y)$ along with an additional spin-orbit interaction assumed to have the Rashba form

$$H \equiv H_0 I + (\alpha/\hbar)(\sigma_x p_y - \sigma_y p_x). \quad (2)$$

We have also investigated the effect of Dresselhaus interactions and edge potentials[14], and find interesting differences like sign reversal, but in this paper we will focus on the Rashba interaction. We study the spin accumulation in this conductor in different transport regimes, from ballistic to diffusive, using the NEGF approach, which has been described elsewhere in detail[12,13]. Here we simply note that elastic incoherent scattering is introduced through a tensor $D$ relating the self-energies to the Green's functions[14] at the same energy $E$

$$\Sigma_s^{in}(i,j;E) = D(i,j,k,l) G^n(k,l;E), \quad (3a)$$

$$\Sigma_s(i,j;E) = D(i,j,k,l) G(k,l;E), \quad (3b)$$

where $i$, $j$, $k$, $l$ can be either spatial or spin indices (Note: $\Sigma^{in} \equiv -i\Sigma^<$, $G^n \equiv -iG^<$, $G \equiv G^R$, $\Sigma \equiv \Sigma^R$). We have shown elsewhere that different choices of the tensor $D$ could allow independent control of phase, momentum and/or spin relaxation [12][16]. In this paper, we will use the uncorrelated point scatterer model for spatial coordinates, which introduces momentum relaxation [17]. However for spin coordinates, we will use two models; one of which introduces spin relaxation ($D(i,j,k,l) = \vec{\sigma}_{ik} \bullet \vec{\sigma}_{lj}$) [16] and another which conserves spin ($D(i,j,k,l) = 1$). By combining these two choices, we obtain desired values for momentum relaxation length $L_m$ and the spin coherence length $L_s$.

---

[1] rgolizad@purdue.edu

[2] datta@purdue.edu

*"Spin-Hall" versus normal Hall effect:* To clarify our model and what we mean by "spin-Hall effect", we start by comparing it with the normal Hall effect for a ballistic rectangular conductor modeled using identical parameters chosen to correspond to the GaAs epilayer used in Ref. 2: $m^*=0.07\,m_e$ ($m_e$: free electron mass), $\alpha = 1.8\,meV-nm$ (as measured [18]), $k_f \cong 10^8 m^{-1}$ corresponding to an electron density of $n = k_f^3/3\pi^2 \cong 3\times 10^{16} cm^{-3}$ and $E_f = 5meV$. Fig.1(a) shows the net z-directed spin density per unit applied bias (with no magnetic field present). Fig. 1(b) shows the **change** in the electron density per unit applied bias relative to the equilibrium (zero bias) electron density with a fixed magnetic field $B$ (= 0.05T) in the z-direction which is included in Eq.(1) through a vector potential [19]. The oscillations in either case correspond to the Fermi wavenumber and are smoothed out on averaging over a range of energies ~2.5meV (~ $K_B T$, T=30K) as shown. We find that the linear accumulation across the sample is described approximately by ($-W/2 < y < +W/2$)

$$\tilde{n}_z(y,E) \approx (k_\alpha/\pi)^2 (M(E)/E)(-2y/W), \qquad (4a)$$

for spin (z-directed) accumulation in the "spin-Hall effect" and by

$$\tilde{n}(y,E) \approx (eB/\pi\hbar)(M(E)/E)(2y/W), \qquad (4b)$$

for electron accumulation in the normal Hall effect. Here $k_\alpha \equiv m^*\alpha/\hbar^2$ is the spin precession wavevector and $M(E)$ is the number of transverse modes or subbands which can be estimated from the expression $M = 2kW/\pi$, $k$ being the electron wavenumber ($=\sqrt{2m^*E}/\hbar$). Eqs. (4a,b) describe our numerical results for a ballistic conductor fairly well and can be justified from a Landauer approach as described below. We can relate the electron accumulation to a Hall voltage $V_H$ through the 2-D density of states to write $V_H = V_A [\tilde{n}]_{max} /(m^*/\pi\hbar^2)$. Noting that the current $I$ is related to the applied bias $V_A$ through $I = (e/h)M(E)V_A$. It is straightforward to show that the Hall resistance deduced from Eq. (4b) agrees with the standard result ($B/en_s$). When we use our NEGF-based approach to introduce incoherent scattering processes, we find that the spin (electron) accumulation is localized within a few mean free paths $L_m$ (not the spin coherence length $L_s$) near the edges. The magnitude is still well described by Eqs. (4a,b); provided we replace the applied voltage with an effective voltage equal to the voltage drop across a mean free path, as explained later in this paper.

The maximum spin accumulation at the edge can be written from Eq. (4a) as

$$[\tilde{n}_z]_{max} \approx (m^*/\pi\hbar^2)(k_\alpha/k_f)(k_\alpha W/\pi). \qquad (5a)$$

For values of $k_\alpha W > 1$, however, the spin accumulation does not increase quadratically with $k_\alpha$ as predicted by Eq. (5a) and is better described by a linear increase ($k_\alpha/k_f < 1$)

$$[\tilde{n}_z]_{max} \approx (m^*/\pi\hbar^2)(k_\alpha/k_f)(1/2\pi). \qquad (5b)$$

More importantly, the spin accumulation is localized near the edges reminiscent of edge states in the quantum Hall regime. Indeed the similarity between Eqs. (4a,b) suggests that for the "spin-Hall effect", the quantity $1/k_\alpha$ plays a role analogous to the magnetic length $\sqrt{\hbar/eB}$ in the normal Hall effect and it seems natural to expect a qualitative change when the quantity $k_\alpha W >> 1$. For the examples presented in this paper, however, $k_\alpha W \leq 1$.

*Landauer approach:* Although our numerical results are obtained from an NEGF-based model, it is instructive to note that Eqs. (4a,b) for a ballistic conductor can be justified from a Landauer viewpoint. We can show that for current flow along $x$ in a clean intrinsic 2-D conductor, the accumulated spin density $n_z(y)$ along the transverse direction $y$ can be written as

$$n_z(y) = \int dE \tilde{n}_z(y,E)[f_1(E) - f_2(E)]. \qquad (6)$$

Note the similarity of Eq. (6) with the standard expression for the current in a ballistic conductor in the Landauer framework [19]

$$I = (e/h)\int dE M(E)[f_1(E) - f_2(E)]. \qquad (7)$$

Eq. (7) is obtained by arguing that for every eigenstate $\exp(+ik_x x)$ there is a degenerate eigenstate $\exp(-ik_x x)$ carrying current in the opposite direction. At equilibrium both eigenstates are equally occupied giving zero net current, but under bias they are occupied from the two contacts with different Fermi functions $f_1$ and $f_2$ giving rise to a net current given by Eq. (7). We can similarly argue that for every eigenstate of H [Eq. (2)] $[p(y)\ q(y)]^T (1/\sqrt{L})\exp(+ik_x x)$ with a net z-directed spin of $|p(y)|^2 - |q(y)|^2$, there is a degenerate eigenstate

$[q(y) \ p(y)]^T (1/\sqrt{L}) \exp(-ik_x x)$ with the opposite spin distribution $|q(y)|^2 - |p(y)|^2$ (*L*: normalization length along *x*-direction). At equilibrium, both eigenstates are occupied and there is no spin accumulation. But under bias, these states are unequally occupied according to different Fermi functions $f_1$ and $f_2$ giving a net spin accumulation described by Eq. (6), with the function

$$\tilde{n}_z(y,E)dE = \sum_{E \leq \varepsilon(n,k_x) \leq E+dE} (1/L) \left[ |p(y)|^2 - |q(y)|^2 \right]_{n,k_x}, \quad (8)$$

obtained by summing the spin accumulation due to each eigenstate having an energy $\varepsilon(n,k_x)$ lying between *E* and *E+dE*. The sign of the spin accumulation alternates between subbands '*n*' but when summed over a large number of modes the function $\tilde{n}_z(y,E)$ is approximated well by the simple expression given in Eq. (4a). Eq. (4b) for the normal Hall effect can also be obtained from the same approach, indeed more simply since the wavefunctions for all eigenstates with $k_x > 0$ are shifted to one edge while those for eigenstates with $k_x < 0$ are shifted to the other edge [Ref. 19, p. 37]. There is thus no alternation in the signs of the terms appearing in the normal Hall version of Eq. (8), making Eq. (4b) relatively straightforward to obtain.

*Diffusive transport:* We can also use the ballistic result in Eq. (5a) to understand spin accumulation in the diffusive regime appropriate for all recent experimental results if we make two important replacements. **Firstly,** the width *W*, which enters Eq. (5a) through $M = 2k_f W/\pi$ should be replaced by an effective width determined by the momentum relaxation length $L_m$

$$W_{eff} = (10 W L_m/(W/2 + 10 L_m)) \sim 20 L_m. \quad (9a)$$

This can be justified by noting that scattering localizes the spin accumulation to $\sim 10 L_m$ near the edges (Fig. 2d). **Secondly,** for a diffusive sample we should use an effective voltage obtained by scaling down the full voltage $V_A$ applied across the length of the conductor *L* by the factor $L_m/(L+L_m)$ (*F*: electric field):

$$eV_{eff} = eV_A L_m/(L+L_m) = eF L L_m/(L+L_m) \quad (9b)$$

This can be justified by noting that in a diffusive sample the separation in the electrochemical potentials for +*k* and –*k* states is $L_m/(L+L_m)$ times the applied voltage [Ref. 19, p. 75]. Since the current is also reduced by the same factor this has no effect on the spin accumulation per unit current.

*Numerical results:* Fig. 2(a) shows one of the structures studied experimentally in Ref. 2. At negative values of *x*, the only edges are at the two ends, while for positive values of *x*, there is a notch in the middle with two additional edges.

Fig. 2(c) shows the spin accumulation measured using Kerr rotation spectroscopy adapted from Ref. 2. The calculated spin density in Fig. 2(d) looks very similar in shape to the experimental results (including the "anticipatory" effect ahead of the notch) and also agrees with the ***sign*** of the effect. Note that while the experimental structure has dimensions ~50μ*m*, the structure studied theoretically has dimensions ~350*nm*. However, in the theoretical model we have deliberately introduced a large amount of scattering such that the mean free path, $L_m$~7nm, making the ratio $L_m/W$ comparable to the experimental structure. Eq. (5a) provides a fairly accurate description of all our numerical results from the ballistic to the diffusive regime; provided we use $W_{eff}$ from Eq.(9a) and multiply by $eV_{eff}$ from Eq.(9b). The reduction in spin accumulation in the diffusive case [Fig. 2(d)] from the ballistic case [Fig. 2(b)] by a factor of ~0.02 can be understood as a product of two factors: reduction in $W_{eff}$ by 0.28 [from Eq. (9a)] and reduction in $eV_{eff}$ by 0.065 [from Eq. (9b)].

Note that the experimental results for the spin density were measured optically at the surface of a GaAs layer, 2μ*m* thick, while our theoretical results are obtained for a 2-D conductor whose parameters (*m**, $k_f$ and $\alpha$) are chosen to reflect those for the experimental structure, as mentioned earlier. We are using a Rashba coefficient $\alpha$=1.8*meV-nm* appropriate for confined GaAs layers, assuming that this is representative of the conditions at the surface of a bulk sample. This seems justifiable since the numerical simulations show that the effect of an edge extends a distance of ~10$L_m$ and the experimental layer thickness is only a fraction of this distance.

In order to compare the magnitude of calculated spin accumulation with experiment, we note that in the experiment *F*=10*meV/μm*, $L_m$≈0.3μ*m*<<*L*≈300μ*m* (assuming a mobility of 3-4*m²/V-s* [20]) so that $eV_{eff}$=3*meV* [Eq. (9b)], while $W_{eff}$≈6μm [Eq. (9a)]. Using these factors to scale the numerical result from Fig. 2(b) we obtain ~40 spins per μ*m²* or 80 spins per μ*m³* (since effective thickness ~*0.5μm*), which is comparable or larger than the spin accumulation observed experimentally, suggesting that the experiments in GaAs can be understood without invoking additional effects. However, the spin accumulation in experiments on ZnSe [21] has a reversed sign and requires a different mechanism.

Two more interesting structures (see Fig. 3 and 4) studied in Ref. 2 show the diffusion of the accumulated spin away from the main current path into protruding sidearms and these are also in good agreement with our numerical model, especially when the spin coherence length $L_s$ is adjusted to have the same ratio $L_s/W$ as the experimental structures [Figs. (3-4)]. Figs. 3, 4 suggest that spatial extend of spin accumulation is independent of $L_s$ in the main channel (*y*=0), but depends strongly on $L_s$ in the sidearm (*y* > 300).

*Concluding remarks:* We have presented numerical results showing striking similarity to recent experimental observations of the "spin-Hall effect" in GaAs [2] and provided approximate expressions [Eqs. (5,9)] that describe the numerical results fairly well both in the ballistic and diffusive limits. Our model predicts that a spin accumulation proportional to the current should be observed even in clean ballistic samples. Furthermore we note that for a specified carrier density (and hence $k_f$), Eq. (5b) predicts a spin accumulation $\sim \alpha m^{*2}$, suggesting that a weak spin-orbit coupling could be compensated for by a large effective mass $m^*$, as in say Silicon ($\alpha \cong 0.4 mev-nm$ [22]).

This work was supported by the MARCO focus center on Materials, Structures and Devices and the NSF Network for Computational Nanotechnology (NCN).

[1] For recent reviews, see H. Engel, E. I. Rashba, B. I. Halperin, cond-mat/0603306, Sept 2006; J. Scheliman, cond-mat/0602330, Apr 2006.
[2] V. Sih, W. H. Lau, R. C. Myers, V. R. Horowitz, A. C. Gossard, and D. D. Awschalom, Phys. Rev. Lett. **97**, 096605 (2006).
[3] J. Wunderlich, B. Kaestner, J. Sinova, and T. Jungwirth, Phys. Rev. Lett. **94**, 047204 (2005).
[4] Y. K. Kato, R. C. Myers, A. C. Gossard, and D. D. Awschalom, Science **306**, 1910 (2004).
[5] S. O. Valenzuela and M. Tinkham, Nature **442**, 176 (2006).
[6] V. M. Galitski, A. A. Burkov, and S. Das Sarma, Phys. Rev. B **74**, 115331 (2006).
[7] J. Sinova, D. Culcer, Q. Niu, N. A. Sinitsyn, and A. H. MacDonald, Phys. Rev. Lett. **92**, 126603 (2004).
[8] S. Murakami, N. Nagaosa, and S. C. Zhang, Science **301**, 1348 (2003).
[9] Wang-Kong Tse, J. Fabian, I. Zutic, and S. Das Sarma, Phys. Rev. B **72**, 241303 (2005).
[10] W. Ren, Z. Qiao, J. Wang, Q. Sun, and H. Guo, Phys. Rev. Lett. **97**, 066603 (2006).
[11] J. H. Bardarson, I. Adagideli, and Ph. Jacquod, cond-mat/0610109, Oct 2006.
[12] R. Golizadeh-Mojarad and S. Datta, Phys. Rev. B (Rapid Comm.) **75**, 081301 (2007).
[13] S. Datta, *Quantum Transport: Atom to Transistor* (Cambridge University Press, Cambridge, 2005).
[14] K. Suzuki and S. Kurihara, cond-mat/0611013, Nov 2006.
[15] G. D. Mahan, Phys. Rep. **145**, 251 (1987).
[16] S. Datta, Proc. Of Inter. School of Phys "Enrico Fermi", Italy, (2004).
[17] S. Datta, J. Phys.: Condensed Matter **2**, 8023 (1990).
[18] V. Sih, R. C. Myers, Y. Kato, W. H. Lau, A. C. Gossard, and D. D. Awschalom, Nature Phys. **1**, 31 (2005).
[19] S. Datta, *Electronic Transport in Mesoscopic Systems* (Cambridge University Press, Cambridge, 1997).
[20] G. E. Stillman, C. M. Wolfe, and J. O. Dimmock, J. Phys. Chem. Solids **31**, 1199 (1970).
[21] N. P. Stern, S. Ghosh, G. Xiang, M. Zhu, N. Samarth, and D. D. Awschalom, Phys. Rev. Lett. **97**, 126603 (2006).
[22] C. Tahan and R. Joynt, Phys. Rev. B **71**, 075315 (2005).

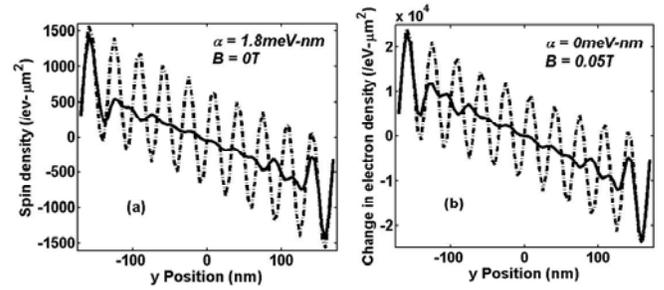

FIG. 1. Comparison of (a) the "spin-Hall effect" with (b) the normal Hall effect in a rectangular ballistic sample. Both calculations are based on the NEGF-based model. Dashed line: single energy calculation; solid line: average over the energy range of $E_f \pm K_B T$ calculation ($E_f$=5meV, T=30K).

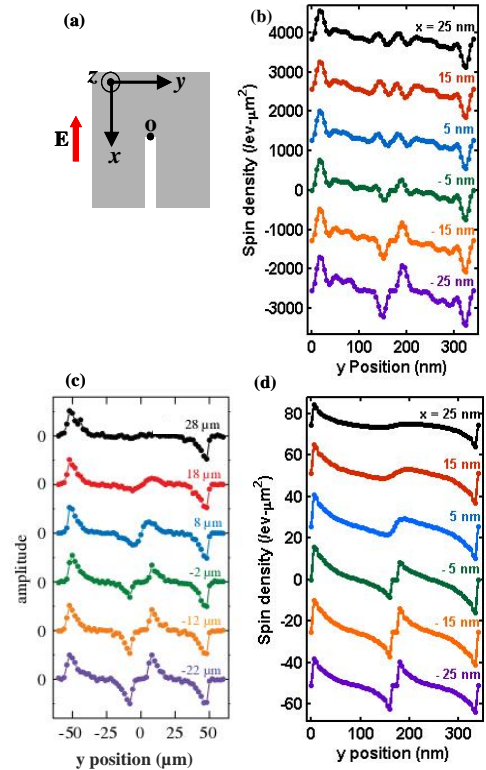

FIG. 2. (a) Schematic shows a channel splitting into two smaller channels. *O* indicates the origin. (b) NEGF-based calculated spin density in ballistic regime averaged over energy range of $E_f \pm K_B T$ as a function of transverse position *y* for different longitudeinal position *x* ($E_f$=5meV, T=30K). (c) Measured Kerr rotation amplitude as a function of transverse position *y* for different longitudinal positions *x* after Ref. 2. (d) NEGF-based calculated spin density in the diffusive regime as a function of transverse position *y* for different longitudinal positions *x* ($L_m \approx 7nm$).

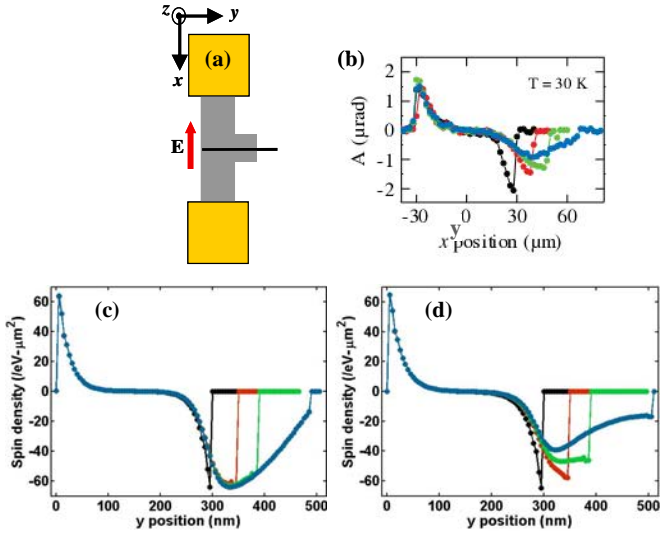

FIG. 3. (a) Schematic device geometry. (b) Kerr rotation as a function of position $y$ for different sidearm lengths after Ref. 2. NEGF-based calculated spin density as a function of position $y$ in diffusive regime for different sidearm lengths: (c) ($L_s \approx \infty$, $L_m = 7nm$). (d) ($L_s = 800nm$, $L_m = 7nm$).

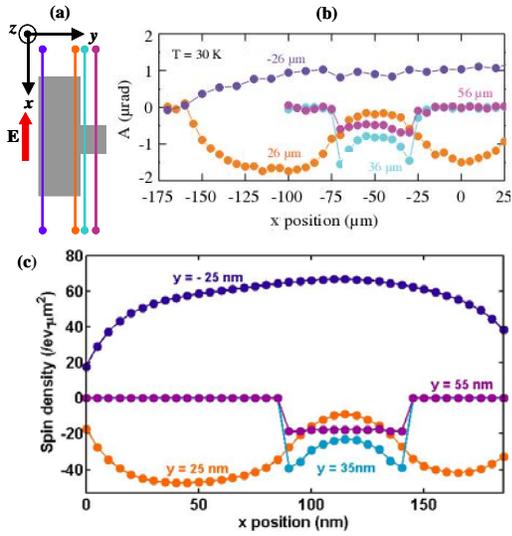

FIG. 4. (a) Schematic showing sample dimension. (b) Amplitude of Kerr rotation measured as a function of longitude position $x$ for different transversal position $y$ in the device after Ref. 2. (c) Calculated spin density in the diffusive regime as a function of longitude position $x$ for different transversal position $y$ ($L_s = 800nm$, $L_m = 7nm$).